\documentclass[runningheads]{llncs}
\usepackage[utf8]{inputenc}
\usepackage{textcomp}
\usepackage{array}

\usepackage{float}
\usepackage{todonotes}
\usepackage{listings}
\usepackage{multirow}
\usepackage{color,soul}
\usepackage{hyperref}
\usepackage{todonotes}
\usepackage{svg}
\usepackage{amsmath}
\usepackage{graphicx}

\usepackage{url}

\begin{document}
\raggedbottom
\title{Perception-Aware Bias Detection for Query Suggestions}

\author{Fabian Haak  \and
Philipp Schaer }
\authorrunning{F. Haak, P. Schaer}

\institute{TH Köln - University of Applied Sciences, Germany\\ 
\email{fabian.haak@th-koeln.de}}
\maketitle              % typeset the header of the contribution

\begin{abstract}
Bias in web search has been in the spotlight of bias detection research for quite a while. At the same time, little attention has been paid to query suggestions in this regard. Awareness of the problem of biased query suggestions has been raised. Likewise, there is a rising need for automatic bias detection approaches. 
This paper adds on the bias detection pipeline for bias detection in query suggestions of person-related search developed by Bonart et al. \cite{Bonart_2019a}. The sparseness and lack of contextual metadata of query suggestions make them a difficult subject for bias detection. Furthermore, query suggestions are perceived very briefly and subliminally. To overcome these issues, perception-aware metrics are introduced.
Consequently, the enhanced pipeline is able to better detect systematic topical bias in search engine query suggestions for person-related searches. The results of an analysis performed with the developed pipeline confirm this assumption. Due to the perception-aware bias detection metrics, findings produced by the pipeline can be assumed to reflect bias that users would discern.

\end{abstract}

\section{Introduction}

Fairness in online search and bias detection are important topics in information retrieval research. Consequently, there are many approaches for detecting bias in search results. Little research and few methodological approaches exist for bias detection in query suggestions. Query suggestions are an important aspect of online information retrieval via search engines and significantly impact what people search for \cite{Niu_2014}. Due to the sparseness of query suggestions (no author, no sources, no publishing platform, less text) and context-dependency, bias detection of query suggestions is less straight forward than bias detection of search results \cite{Robertson_2019}. Unless a person performing online search does not have a clear information need, very little attention is paid to the query suggestions. Because of the brief exposure, search engine users perceive query suggestions distinctly, even though certain effects like the diminishing attention paid to elements further down the list still applies \cite{dean_2019,hofmann_2014}. 
Summarizing these findings, we are left with two research questions to focus on as we develop a bias detection pipeline:

\begin{itemize}
\item RQ1: To what extent can bias towards metadata on the persons searched (e.g., gender, age, party membership) be uncovered in query suggestions to person-related searches using perception-aware metrics?
\item RQ2: How do perception-aware metrics perform compared to simpler metrics in detecting bias in query suggestions to person-related searches?
\end{itemize}

To answer these research questions, we extend the bias identification pipeline developed by Bonart et al. by introducing perception-aware metrics. Doing so, the pipeline identifies bias in a more realistic manner. This should in turn result in a better indication of present bias. By testing the updated pipeline on the same data, the results produced by both pipelines are directly comparable.

\section{Related Work}
Findings of various studies show that search engines such as Google are seen as a trustworthy source of information on many topics, including political information \cite{ray_2020,edelman_2020}. According to Houle, search engines significantly impact political opinion formation \cite{Houle_2015}. The trust in search engines is problematic because their results are prone to be biased. This might be due to bias induced by algorithms \cite{Introna_2000}, or by representing bias inherent in mental models: Although not at all true by definition of these words, a doctor is usually assumed to be male, while a nurse is typically expected to be female \cite{bolukbasi2016}. Similar biased patterns exist plentiful, and because search engines show the information inherent in data, bias is as omnipresent in search results as it is in people's minds and natural language texts \cite{Noble_2018,DBLP:Dev_2019}. Even Google acknowledges this in a white paper, noting its awareness of disinformation and biased content presented by its search engine \cite{google_2019}. 
Biased search results have been widely discussed and researched. Kulshrestha et al. investigated and compared bias in Google search results and search on Twitter \cite{Kulshrestha_2018}. Their bias detection methodology relies on peripheral information, such as the author, publishing and distribution platforms, as well as information gathered from these sources \cite{Kulshrestha_2018}.

Aside from search results, \textit{query suggestions} play a key role in what people search for \cite{Niu_2014}. There are many types of query suggestions, such as query expansions, auto completions, query predictions, or query refinements \cite{Ooi_2015,DBLP:journals/ftir/CaiR16}. We use the term query suggestion as an umbrella term for all facets of the previously mentioned terms to describe the list of suggested search queries returned by search engines for an input query or search term. Although it is unclear how exactly they are formed, there is no doubt that query suggestions are derived from what other users search for in a location and language \cite{wang_2018}. Also, they can be manipulated \cite{wang_2018}. Query suggestions are assumed to be as bias-laden as search results, with a study by Olteanu et al. illustrating how diverse and hard to detect the forms of problematic and biased search results are \cite{olteanu_2020}. The difficulty in detecting bias in query suggestions lies in their sparseness. Without context, they neither offer author nor source information, nor is input bias available to judge their ranking. Furthermore, bias in query suggestions is often context-dependent and not derivable from the terms themselves \cite{olteanu_2020}. For these reasons, approaches like the one utilized by Kulshrestha et al. do not work for query suggestions.

To overcome the hurdles of identifying bias in sparse, contextless search query suggestions, Bonart et al. developed a bias identification pipeline for person-related search \cite{Bonart_2019a}. It represents a natural language processing pipeline with three modules: Data acquisition, data preprocessing, and bias analysis (cf. fig.~1). The collection they used to develop and test the pipeline consists of search queries and their corresponding lists of query suggestions, gathered since 2017 in German from Google, DuckDuckGo, and Bing. The search terms consist primarily of names of German politicians.
With the developed pipeline, Bonart et al. tried to find a way to automatically identify \textit{systematic topical biases} towards certain groups of people with shared meta-attributes (e.g., gender, age, party membership) \cite{Bonart_2019a}. Topical bias describes content bias as misrepresented information in the documents themselves \cite{Pitoura_2018}. Concerning the meta-attributes, systematic topical bias refers to differences in the distribution of topics in query suggestions of groups of politicians of different meta-attribute characteristics (e.g., male/ female). Due to their aforementioned sparseness, this type of bias is the easiest to detect in query suggestions.

\begin{figure}[t]
\label{fig:bias_frame}
\begin{center}
\includegraphics[width=\textwidth]{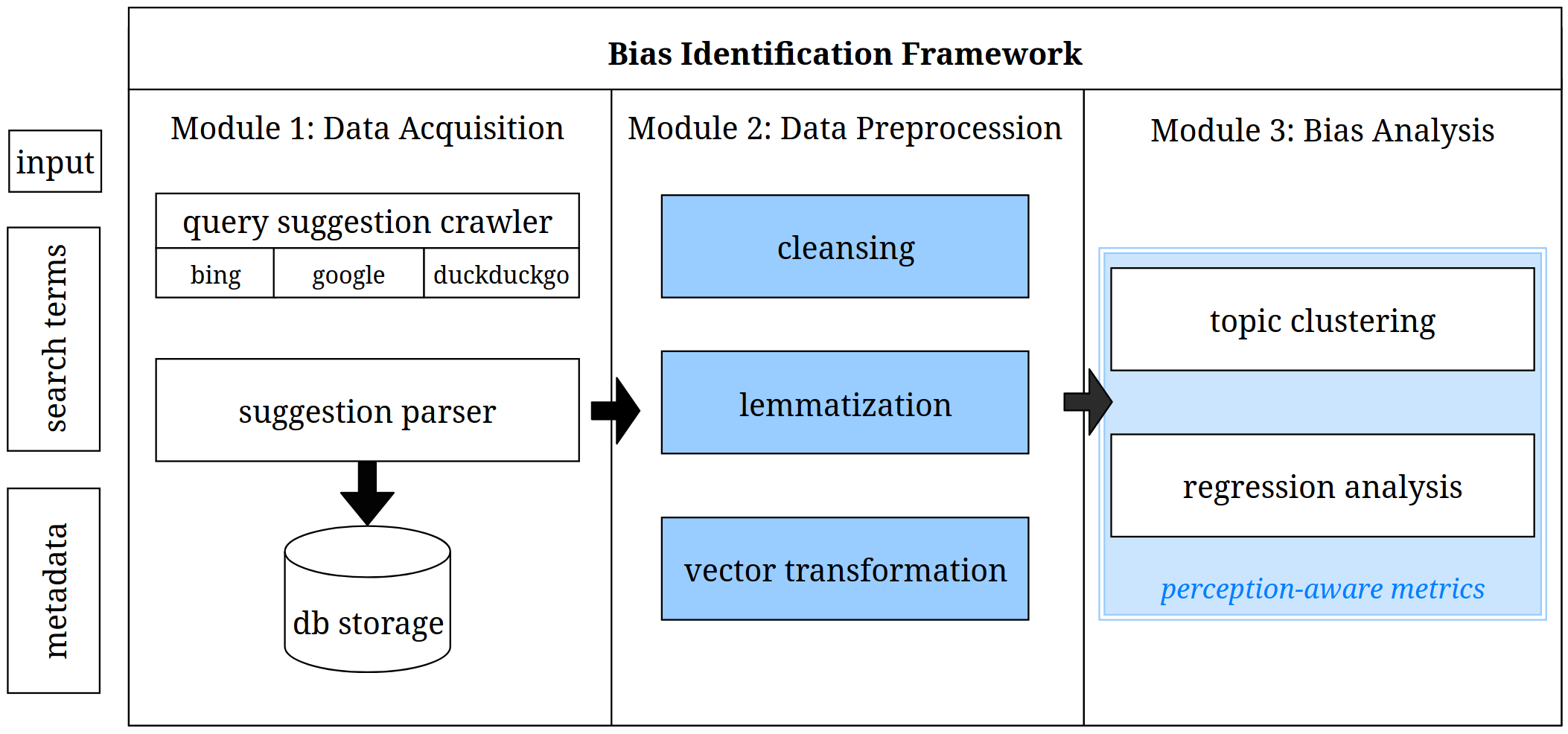}
\end{center}
\caption{Bias detection pipeline for person-related query suggestions developed by Bonart et al. \cite{Bonart_2019a}. The highlighted steps have been modified.}
\end{figure}

The metric Bonart et al. used for detecting bias is the number of unique topical cluster terms in the query suggestions for each search term (cf. section~\ref{methodology}). Bias is then identified by comparing the differences between certain groups of search terms. The results produced by the bias identification pipeline only revealed minor indicators of bias within the groups of politicians. In addition to the insignificant findings, the metric employed by Bonart et al. does not consider two critical aspects of how query suggestions are perceived: (A) The \textit{frequency of the suggestions} throughout data collection is disregarded. A suggestion that appears once influences the metric as much as a suggestion that appeared many times. (B) The \textit{order of items} in the list of query suggestions is ignored. Although the rank in the list of query suggestions has a strong influence on the visibility of a query suggestion \cite{hofmann_2014}, it was not taken into account. 
Because of these flaws in the metrics, it is doubtful whether the detected bias would have been perceivable by search engine users. Therefore, the bias identification pipeline is reworked and expanded to produce more meaningful results and consider how query suggestions are perceived.

\section{Methodology}\label{methodology}
The following section briefly outlines the bias detection pipeline and highlights changes and additions made to it. 
\paragraph{Data Acquisition.}\label{datAq}
The bias analysis is based on a set of N search terms \(t_i\) with \(i=1,...,N\) which share a set of P meta-attributes \({x_{i,1},...,x_{i,P}}\) \cite{Bonart_2019a}. The dataset consists of a collection of query suggestions returned for these terms, which consist of names of German politicians or politically relevant people from Germany. Twice per day, for all search terms, a specialized web crawler collects query suggestions in German by HTTP request from three search engines' auto-complete APIs: Google, DuckDuckGo and Bing. Requests only contain the input query as well as language and location information. Therefore, no user profiles or search histories can influence the results.

\paragraph{Preprocessing.}
Preprocessing is the first module of the pipeline that was changed significantly. The lemmatizer was changed from pattern.de to the german\_news standard model made available within the spacy module by Honnibal et al. \cite{spacy}. An entity recognition step was introduced and is performed using Spacy, as well. It is performed after lemmatization on all query suggestions that do not consist of a single word. After cleaning and lemmatization, about 20\% of the query suggestions are discarded in the original pipeline because clustering can only be performed on single-word suggestions. By employing an entity recognition system, many suggestions such as ``summer festival'' are condensable to a single term, which can be used in the cluster analysis. Since query suggestions are most probably formed considering entities to not deliver multiple very similar suggestions to the user (cf. \cite{DBLP:journals/ftir/CaiR16}), the suggestions shortened by entity recognition are not expected to change significantly in meaning. The last step in the preprocessing module consists of the unmodified vectorization of the single-term query suggestions using the word2vec module \cite{mikolov2013}. Although there are newer and more elaborate vectorizers, word2vec performed best on the given collection of German terms.

\paragraph{Bias Analysis.}\label{percAw}
The first step in the bias analysis module, topic clustering, has not been changed methodically. The vectorized query suggestions are assigned to topical clusters utilizing a k-means approach.

Most importantly, new metrics have been introduced. These metrics are perception-aware, meaning that they aim to determine bias analog to how a user would perceive it. Due to the sparseness of metadata around the suggestions and the often context-dependent bias, topical bias is the most promising approach to automatically detecting bias. A main characteristic of the perception of query suggestions is the low attention given to them, especially suggestions on lower ranks \cite{hofmann_2014,Mitra_2014}. Therefore, the main factors influencing the exposure of a topic over a time span are the percentage of relevant topical suggestions and their ranks. 

As a first step to derive the new metrics, a matrix is created, from which the metrics are calculated. This matrix contains rows for all search terms \(t_{ir}\) with \(i=1,...,N\) being the identifier of the term and \(r=1,...,10\) signifying the rank in the list of query suggestions for the term. Search terms and suggestion rank form the multi-index structure for the rows of the matrix. These rows are paired with all M preprocessed single-term query suggestions \(s_{j}\) with \(j=1,...,M\), that have been assigned to a cluster. This results in a structure where the frequency for every search term-suggestion combination at every rank is stored. Based on this, the number and percentage of suggestions of each of the topical clusters at each rank can be calculated for each search term.  

The problem of judging the systematic topical bias in the query suggestions approximates the relevance judgment. Relevance judgments usually describe the probability of a document to fulfill a given information need. Likewise, the percentage of suggestions for a certain cluster for a rank describes the probability of a randomly selected single term belonging to that topic. Thus, using a discounted relevance measure to judge the rank- and perception-aware topical affiliation is not far-fetched. Hence, discounted cumulative gain and its normalized variety are introduced as metrics for detecting bias. Both are usually employed to rank the relevance of a list of documents, for example, as returned by a search query \cite{lin_2020}. The metrics put an emphasis on the rank of the judged elements, which is what we want our metric to do. Discounted Cumulative Gain (DCG) is implemented and adopted as a bias metric for query suggestions as \cite{Jarvelin_2002}:
\begin{equation}
    DCG(C_{x},q)=\sum_{i=1}^{10}\frac{2^{P(C_x(i),q)}-1}{log_{2}(i+1)},
\end{equation}
where $DCG(C_{x},q)$ describes the DCG of a term q for cluster x and ${P(C_x(i),q)}$ is the percentage of total appearances of clustered query suggestions at rank $i$ of the list of query suggestions for the term. Instead of relevance, we employ the percentage of suggestions for a topical cluster, which can be interpreted as the topical affiliation for that cluster. Counting appearances of cluster words and using the percentages as key measurements is similar to using graded instead of dichotomous relevance judgments. In essence, instead of a measure of gain, DCG as a metric for topical affiliation describes the average degree of perceived exposure to a topic within the query suggestions of a search term. By revealing differences in the topical affiliation between groups, topical bias towards these groups can be identified and quantified.

The nDCG (normalized Discounted Cumulative Gain) for cluster x of a term is expressed as the DCG divided by the DCG of the percentages ${P(C_x(i),q)}$ sorted in descending order (${IDCG(C_{x},q)}$):
\begin{equation}
    nDCG(C_{x},q)=\frac{DCG(C_{x},q)}{IDCG(C_{x},q)}
\end{equation}
By normalizing, the nDCG expresses how every cluster is distributed over the ranks of the query suggestions of a term, neglecting the overall number of cluster words and other clusters. Thereby, it expresses the average rank at which the suggestions of the topical cluster appear. A high nDCG score means that a topical cluster appears on average in the first ranks of the suggestions. However, it does not indicate how often suggestions of the topic appear over the span of data acquisition. The nDCG could be a useful metric if the lengths of query suggestions vary, when particular clusters or terms are uncommon or when very little data is available. For example, when trying to judge how prominent a term appeared in searches, that coined only in a very brief time in search suggestions (e.g., suggestion ``flu'' with names of different states). These terms do not appear often enough over a year to impact the DCG score, but the nDCG allows for differentiated insight anyway by only emphasizing the rank.

\paragraph{Regression Analysis.}
The metrics describe how the identified topical clusters would manifest in the query suggestions for the search terms. The goal is to identify significant differences between the groups of meta-attributes $x_{i,p}$ (e.g., female, SPD-member) in the perception-aware metrics for each cluster $y_{i,c}$ (e.g, DCG, nDCG). By doing so, topical bias (e.g., towards terms that describe private or family topics) is detectable. To reveal significant differences, multiple linear regression is performed using dichotomous dummy variables for the meta-attributes as independent variables and the perception-aware metrics nDCG and DCG as dependent variables. The model of this regression for topical clusters $c\in1,...,k$  can be expressed as
\begin{equation}
    y_{i,c}=\beta_0+\beta_1x_{i,1}+...+\beta_px_{i,p}+\epsilon_i,
\end{equation}
where $\epsilon_i$ is the independent error term and $i=1,...,N$ are the observation indices. To avoid multicollinearity, one variable per attribute is used as the base category and omitted. 

\section{Results of Bias Analysis of German Politicians Dataset}\label{results}
After describing the main changes to the bias detection pipeline in the previous section, this section explores its effectiveness by performing an analysis using the pipeline on the most recent version of the same dataset of German politicians used to test the first version of the pipeline.
\paragraph{Data Acquisition.}
As mentioned, the dataset consists of German politicians' names as search terms and the corresponding returned query suggestions. Compared to when the bias identification pipeline was first developed, the list was expanded significantly. The number of search terms was raised from 630 to 3047. The additional terms consist of politicians who are currently members of the Bundestag (the federal parliament) but were not in 2017 when the dataset was created. Additionally, political figures who are not members of a German government body were added to the dataset. Some names of politicians with a city name attached to it have also been added (e.g. ``Volker Beck (Köln)''), along with some terms that either do not describe a person or are for some reason misspelled variants of names (e.g. ``spd'' or ``goeringeckardt''). Both types have been filtered out. As meta-attributes, further information was gathered both manually and with a scraper tool, accessing publicly available resources such as personal websites and databases such as Abgeordnetenwatch\footnote{\url{https://www.abgeordnetenwatch.de/}} or Wikidata\footnote{\url{https://www.wikidata.org/wiki/Wikidata:MainPage}}. For each person, the socio-demographic factors age, gender, party affiliation, and federated state of political origin were collected. For 1227 of the search terms, all information was aggregated, doubling the number of search terms. Furthermore, the period of data collection now spans 34 instead of 4 months, vastly expanding the collection. The data set and corresponding results of analyses will be published after the publication of this paper.
The data set includes 33.4 percent female politicians. The average age is 54. Most politicians in the dataset originate from the German federal state of North Rhine-Westphalia (16 percent), while the smallest part originates from the German federal state of Bremen (1.5 percent). The biggest party in the sample is the CDU (``Christian Democratic Union'') with a proportion of about 25 percent. These proportions correspond roughly to the actual distributions of these attributes among German politicians.
\paragraph{Preprocessing.}
The updated preprocessing module with the added entity detection step still has to omit some of the crawled query suggestions. Albeit with a loss of around 18 percent, there is less potential loss of information due to the removal of longer query suggestions. After cleaning, lemmatization, and entity detection, 5405 unique single-word suggestions remained. The vector-transformation algorithm was able to vectorize 3979 of these words. 

\paragraph{Bias Analysis.}
The word embedding vectors for each of the suggestions were used to perform a cluster analysis. A k-means approach was performed with three clusters, as suggested by the employed heuristics. By manually evaluating the clusters, we assigned a label that best describes the topic of each cluster (cf. table~\ref{tab:cluster_overview}). The first cluster includes terms with personal meaning. The second cluster consists mostly of names of cities and places that are of no political significance. The third group contains words with political meaning ranging from topics (e.g. ``drug report'') over other political persons (e.g. ``Guelen'') to cities and counties that are of political significance (e.g. ``Afghanistan'').
\begin{table}[H]
    \centering
    \caption{Examples for terms of the clusters found by performing a k-means clustering approach on the preprocessed single-word query suggestions. Translated from German to English.}
    \begin{tabular}{m{4cm} | m{4cm} | m{4cm} }
      Cluster 1:  & Cluster 2: & Cluster 3: \\ 
      Personal & Cities and Places & Politics and Economics \\ \hline
      losing weight  & Aachen & Airbus\\
      events & Cologne & stocks\\
      hair & Bielefeld & Guelen\\
      … & … & …\\
    \end{tabular}

    \label{tab:cluster_overview}
\end{table}
As described in section~\ref{percAw}, bias metrics are calculated based on the position and frequency of suggestions corresponding to the clusters assigned in the previous step. Before calculating the metrics, search terms with less than 10 cluster words are dropped. This reduces the number of search terms to 2510, 1321 of which have a federated state, 1146 a Party, 1238 a gender, and 1253 an age assigned. 1227 of the politicians have all meta-attributes assigned. 

Table 2 shows the results of the multiple linear regression analysis performed on the DCG and nDCG. The CDU and Baden-Württemberg were chosen as base-categories for the attributes party and federated state. 
For all metrics, the F-test rejected the joint null hypothesis that all coefficients are zero. Therefore, relevant biased patterns towards each of the metrics can be considered. Although there are biased patterns for each cluster, cluster 2 shows notably less. Very few attributes show biased patterns towards suggestions of that topical cluster. This reflects in the amount of variance explained by the models for the clusters. The regression model using the DCG scores could explain 7\%, 1\% and 5\% of the variance for clusters 1, 2 and 3. The nDCG performed very similarly with 5\%, 2\% and 6\%, respectively.  

For cluster 2 (names of cities and countries), only politicians of the CSU (Christian Social Union in Bavaria) and the LINKE (Democratic Socialists' Party) exhibit significantly higher DCG values than the base category. The members of the LINKE also have significantly higher nDCG values. Cluster 2 suggestions, names of places without political significance, appear on average 1.5 ranks higher for LINKE politicians than for other parties. 
The perception-aware metrics show a significant topical gender bias towards the cluster of political and economic-related suggestions. The results show significantly ($P<0.01$) lower average DCG scores (cluster 3: male 0.7, female 0.49, cf. figure~2) for suggestions of cluster 3 if the search term is a female person. This also shows in the corresponding nDCG values. With a coefficient of roughly -0.1 (nDCG scores for cluster 3: male 0.46, female 0.36, cf. figure~2), query suggestions with political topics appear on average one rank lower if the searched person is female. 
Age was identified as a biased factor for both cluster 1 and cluster 3. The older the politician, the more politics- and the less personal-related are the query suggestions. Figure 2 shows the mean scores for politicians over and under 40 years of age. The DCG score for cluster 1 is significantly higher for younger politicians, while for cluster 3, the opposite is the case. This also reflects in the regression results. We found some significant bias for both metrics within the political parties and the federated states towards suggestions of the cluster of political terms as well as the cluster of private terms (cf. table~\ref{tab:regression_results}).

\begin{figure}[t]
\label{fig:groups}
\begin{center}
\includegraphics[width=\textwidth]{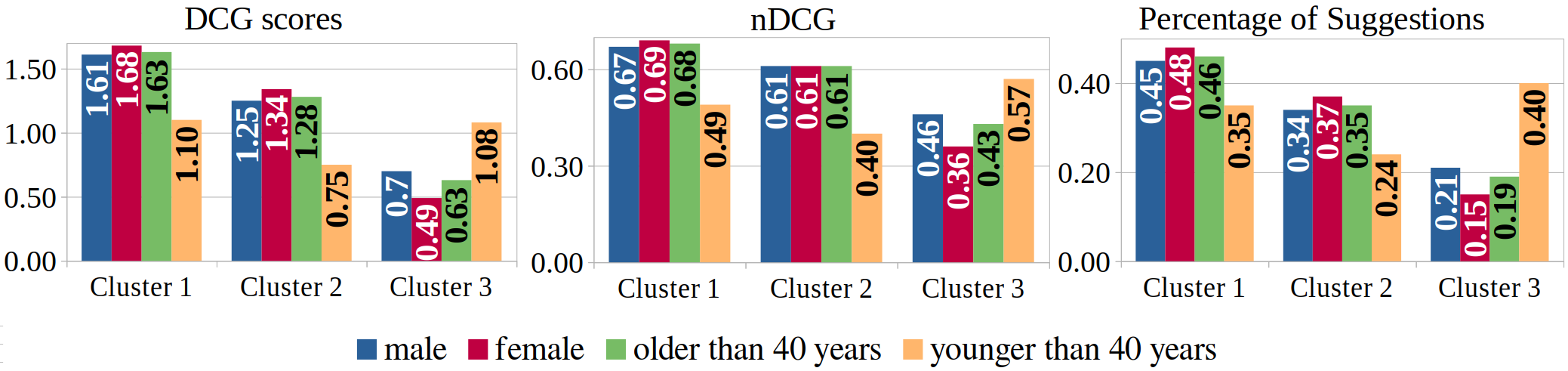}
\end{center}
\caption{DCG and nDCG scores as well as total appearance percentages for gender and age meta-attributes. The dataset includes 818 male and 420 female politicians, 1096 older than or exactly 40 years old and 1414 younger than 40 years.}
\end{figure}

\begin{table}[t]
\label{tab_results}
\caption{Results of the regression analysis for nDCG and DCG scores for each of the clusters. Shown are the coefficients B along with the significance value of the test for coefficients P, for all metric-attribute combinations. The F-test score for overall significance and the adjusted $R^2$ measure ${R^2}_c$ can be found in the row labeled "Model". All values are rounded, significant results ($P<0.05$) are highlighted.}

\resizebox{\textwidth}{!}{%

    \begin{tabular}{llclclclclclc}
\multicolumn{1}{l|}{}                         & \multicolumn{2}{c|}{nDCG\_1}                                           & \multicolumn{2}{c|}{nDCG\_2}                                           & \multicolumn{2}{c|}{nCDG\_3}                                           & \multicolumn{2}{c|}{DCG\_1}                                            & \multicolumn{2}{c|}{DCG\_2}                                            & \multicolumn{2}{c|}{DCG\_3}                                            \\
\multicolumn{1}{l|}{}                         & \multicolumn{1}{c}{B}             & \multicolumn{1}{c|}{P}             & \multicolumn{1}{c}{B}             & \multicolumn{1}{c|}{P}             & \multicolumn{1}{c}{B}             & \multicolumn{1}{c|}{P}             & \multicolumn{1}{c}{B}             & \multicolumn{1}{c|}{P}             & \multicolumn{1}{c}{B}             & \multicolumn{1}{c|}{P}             & \multicolumn{1}{c}{B}             & \multicolumn{1}{c|}{P}             \\ \hline
\multicolumn{1}{l|}{(constant)}               & \textbf{0.80}                     & \multicolumn{1}{c|}{\textbf{0.00}} & \textbf{0.60}                     & \multicolumn{1}{c|}{\textbf{0.00}} & \textbf{0.38}                     & \multicolumn{1}{c|}{\textbf{0.00}} & \textbf{2.24}                     & \multicolumn{1}{c|}{\textbf{0.00}} & \textbf{1.28}                     & \multicolumn{1}{c|}{\textbf{0.00}} & \textbf{0.43}                     & \multicolumn{1}{c|}{\textbf{0.00}} \\ \hline
\multicolumn{1}{l|}{female}                   & 0.02                              & \multicolumn{1}{c|}{0.31}          & -0.01                             & \multicolumn{1}{c|}{0.74}          & \textbf{-0.09}                    & \multicolumn{1}{c|}{\textbf{0.00}} & 0.05                              & \multicolumn{1}{c|}{0.38}          & 0.08                              & \multicolumn{1}{c|}{0.19}          & \textbf{-0.20}                    & \multicolumn{1}{c|}{\textbf{0.00}} \\ \hline
\multicolumn{1}{l|}{age (groups of 10 years)} & \textbf{-0.03}                    & \multicolumn{1}{c|}{\textbf{0.00}} & 0.00                              & \multicolumn{1}{c|}{0.96}          & \textbf{0.02}                     & \multicolumn{1}{c|}{\textbf{0.01}} & \textbf{-0.14}                    & \multicolumn{1}{c|}{\textbf{0.00}} & 0.00                              & \multicolumn{1}{c|}{0.94}          & \textbf{0.07}                     & \multicolumn{1}{c|}{\textbf{0.00}} \\ \hline
Baden-Württemberg                             & \multicolumn{12}{c}{reference category}                                                                                                                                                                                                                                                                                                                                                                                                             \\
\multicolumn{1}{l|}{Bayern}                   & 0.05                              & \multicolumn{1}{c|}{0.18}          & 0.02                              & \multicolumn{1}{c|}{0.60}          & 0.02                              & \multicolumn{1}{c|}{0.71}          & 0.25                              & \multicolumn{1}{c|}{0.05}          & -0.12                             & \multicolumn{1}{c|}{0.36}          & 0.02                              & \multicolumn{1}{c|}{0.85}          \\
\multicolumn{1}{l|}{Berlin}                   & 0.02                              & \multicolumn{1}{c|}{0.56}          & -0.04                             & \multicolumn{1}{c|}{0.33}          & \textbf{-0.13}                    & \multicolumn{1}{c|}{\textbf{0.00}} & 0.04                              & \multicolumn{1}{c|}{0.75}          & -0.09                             & \multicolumn{1}{c|}{0.47}          & -0.03                             & \multicolumn{1}{c|}{0.78}          \\
\multicolumn{1}{l|}{Brandenburg}              & 0.08                              & \multicolumn{1}{c|}{0.10}          & 0.00                              & \multicolumn{1}{c|}{0.99}          & -0.08                             & \multicolumn{1}{c|}{0.18}          & 0.21                              & \multicolumn{1}{c|}{0.26}          & 0.03                              & \multicolumn{1}{c|}{0.86}          & -0.17                             & \multicolumn{1}{c|}{0.23}          \\
\multicolumn{1}{l|}{Bremen}                   & -0.06                             & \multicolumn{1}{c|}{0.36}          & -0.16                             & \multicolumn{1}{c|}{0.05}          & \textbf{0.34}                     & \multicolumn{1}{c|}{\textbf{0.00}} & -0.25                             & \multicolumn{1}{c|}{0.34}          & -0.34                             & \multicolumn{1}{c|}{0.19}          & \textbf{0.46}                     & \multicolumn{1}{c|}{\textbf{0.03}} \\
\multicolumn{1}{l|}{Hamburg}                  & 0.09                              & \multicolumn{1}{c|}{0.06}          & 0.01                              & \multicolumn{1}{c|}{0.80}          & -0.11                             & \multicolumn{1}{c|}{0.05}          & 0.06                              & \multicolumn{1}{c|}{0.75}          & 0.04                              & \multicolumn{1}{c|}{0.80}          & -0.09                             & \multicolumn{1}{c|}{0.51}          \\
\multicolumn{1}{l|}{Hessen}                   & -0.03                             & \multicolumn{1}{c|}{0.44}          & 0.02                              & \multicolumn{1}{c|}{0.63}          & 0.03                              & \multicolumn{1}{c|}{0.45}          & -0.05                             & \multicolumn{1}{c|}{0.69}          & -0.07                             & \multicolumn{1}{c|}{0.56}          & 0.16                              & \multicolumn{1}{c|}{0.11}          \\
\multicolumn{1}{l|}{Mecklenburg-Vorpommern}   & -0.05                             & \multicolumn{1}{c|}{0.37}          & 0.03                              & \multicolumn{1}{c|}{0.62}          & -0.03                             & \multicolumn{1}{c|}{0.61}          & -0.23                             & \multicolumn{1}{c|}{0.25}          & 0.10                              & \multicolumn{1}{c|}{0.60}          & -0.17                             & \multicolumn{1}{c|}{0.27}          \\
\multicolumn{1}{l|}{Niedersachsen}            & 0.03                              & \multicolumn{1}{c|}{0.28}          & -0.02                             & \multicolumn{1}{c|}{0.57}          & 0.00                              & \multicolumn{1}{c|}{0.91}          & 0.04                              & \multicolumn{1}{c|}{0.71}          & -0.19                             & \multicolumn{1}{c|}{0.10}          & -0.02                             & \multicolumn{1}{c|}{0.87}          \\
\multicolumn{1}{l|}{Nordrhein-Westfalen}      & 0.05                              & \multicolumn{1}{c|}{0.07}          & -0.01                             & \multicolumn{1}{c|}{0.80}          & 0.02                              & \multicolumn{1}{c|}{0.65}          & \textbf{0.22}                     & \multicolumn{1}{c|}{\textbf{0.03}} & -0.07                             & \multicolumn{1}{c|}{0.50}          & -0.06                             & \multicolumn{1}{c|}{0.43}          \\
\multicolumn{1}{l|}{Rheinland-Pfalz}          & 0.06                              & \multicolumn{1}{c|}{0.16}          & 0.01                              & \multicolumn{1}{c|}{0.78}          & 0.03                              & \multicolumn{1}{c|}{0.63}          & 0.11                              & \multicolumn{1}{c|}{0.48}          & 0.06                              & \multicolumn{1}{c|}{0.70}          & -0.03                             & \multicolumn{1}{c|}{0.81}          \\
\multicolumn{1}{l|}{Saarland}                 & \textbf{0.13}                     & \multicolumn{1}{c|}{\textbf{0.04}} & -0.04                             & \multicolumn{1}{c|}{0.58}          & -0.09                             & \multicolumn{1}{c|}{0.24}          & 0.16                              & \multicolumn{1}{c|}{0.49}          & 0.00                              & \multicolumn{1}{c|}{0.99}          & -0.25                             & \multicolumn{1}{c|}{0.16}          \\
\multicolumn{1}{l|}{Sachsen}                  & 0.01                              & \multicolumn{1}{c|}{0.77}          & 0.01                              & \multicolumn{1}{c|}{0.82}          & 0.02                              & \multicolumn{1}{c|}{0.71}          & -0.03                             & \multicolumn{1}{c|}{0.82}          & 0.01                              & \multicolumn{1}{c|}{0.94}          & 0.05                              & \multicolumn{1}{c|}{0.60}          \\
\multicolumn{1}{l|}{Schleswig-Holstein}       & -0.01                             & \multicolumn{1}{c|}{0.78}          & -0.01                             & \multicolumn{1}{c|}{0.80}          & -0.01                             & \multicolumn{1}{c|}{0.88}          & -0.08                             & \multicolumn{1}{c|}{0.60}          & 0.03                              & \multicolumn{1}{c|}{0.83}          & -0.16                             & \multicolumn{1}{c|}{0.18}          \\
\multicolumn{1}{l|}{Thüringen}                & 0.06                              & \multicolumn{1}{c|}{0.18}          & -0.03                             & \multicolumn{1}{c|}{0.55}          & \textbf{0.12}                     & \multicolumn{1}{c|}{\textbf{0.04}} & 0.04                              & \multicolumn{1}{c|}{0.82}          & -0.13                             & \multicolumn{1}{c|}{0.43}          & 0.15                              & \multicolumn{1}{c|}{0.26}          \\
\multicolumn{1}{l|}{Sachsen-Anhalt}           & 0.01                              & \multicolumn{1}{c|}{0.85}          & 0.00                              & \multicolumn{1}{c|}{0.96}          & 0.01                              & \multicolumn{1}{c|}{0.94}          & -0.09                             & \multicolumn{1}{c|}{0.66}          & -0.01                             & \multicolumn{1}{c|}{0.97}          & 0.02                              & \multicolumn{1}{c|}{0.90}          \\ \hline
CDU                                           & \multicolumn{12}{c}{reference category}                                                                                                                                                                                                                                                                                                                                                                                                             \\
\multicolumn{1}{l|}{SPD}                      & 0.02                              & \multicolumn{1}{c|}{0.36}          & 0.03                              & \multicolumn{1}{c|}{0.29}          & -0.05                             & \multicolumn{1}{c|}{0.07}          & 0.07                              & \multicolumn{1}{c|}{0.36}          & 0.09                              & \multicolumn{1}{c|}{0.22}          & -0.12                             & \multicolumn{1}{c|}{0.05}          \\
\multicolumn{1}{l|}{CSU}                      & -0.03                             & \multicolumn{1}{c|}{0.52}          & 0.08                              & \multicolumn{1}{c|}{0.11}          & -0.05                             & \multicolumn{1}{c|}{0.37}          & -0.22                             & \multicolumn{1}{c|}{0.19}          & \textbf{0.53}                     & \multicolumn{1}{c|}{\textbf{0.00}} & -0.21                             & \multicolumn{1}{c|}{0.11}          \\
\multicolumn{1}{l|}{other parties}            & 0.06                              & \multicolumn{1}{c|}{0.31}          & 0.02                              & \multicolumn{1}{c|}{0.71}          & 0.03                              & \multicolumn{1}{c|}{0.73}          & 0.22                              & \multicolumn{1}{c|}{0.30}          & -0.02                             & \multicolumn{1}{c|}{0.94}          & -0.15                             & \multicolumn{1}{c|}{0.36}          \\
\multicolumn{1}{l|}{AFD}                      & \textbf{0.11}                     & \multicolumn{1}{c|}{\textbf{0.00}} & -0.01                             & \multicolumn{1}{c|}{0.83}          & \textbf{-0.13}                    & \multicolumn{1}{c|}{\textbf{0.00}} & \textbf{0.65}                     & \multicolumn{1}{c|}{\textbf{0.00}} & -0.12                             & \multicolumn{1}{c|}{0.25}          & \textbf{-0.41}                    & \multicolumn{1}{c|}{\textbf{0.00}} \\
\multicolumn{1}{l|}{LINKE}                    & -0.01                             & \multicolumn{1}{c|}{0.78}          & \textbf{0.17}                     & \multicolumn{1}{c|}{\textbf{0.00}} & \textbf{-0.08}                    & \multicolumn{1}{c|}{\textbf{0.02}} & -0.04                             & \multicolumn{1}{c|}{0.69}          & \textbf{0.29}                     & \multicolumn{1}{c|}{\textbf{0.01}} & \textbf{-0.23}                    & \multicolumn{1}{c|}{\textbf{0.01}} \\
\multicolumn{1}{l|}{FDP}                      & 0.01                              & \multicolumn{1}{c|}{0.68}          & -0.03                             & \multicolumn{1}{c|}{0.37}          & -0.07                             & \multicolumn{1}{c|}{0.07}          & 0.08                              & \multicolumn{1}{c|}{0.47}          & -0.08                             & \multicolumn{1}{c|}{0.45}          & \textbf{-0.18}                    & \multicolumn{1}{c|}{\textbf{0.04}} \\
\multicolumn{1}{l|}{GRÜNE}                    & \textbf{0.11}                     & \multicolumn{1}{c|}{\textbf{0.00}} & -0.02                             & \multicolumn{1}{c|}{0.54}          & \textbf{-0.10}                    & \multicolumn{1}{c|}{\textbf{0.00}} & \textbf{0.42}                     & \multicolumn{1}{c|}{\textbf{0.00}} & -0.10                             & \multicolumn{1}{c|}{0.28}          & \textbf{-0.23}                    & \multicolumn{1}{c|}{\textbf{0.00}} \\ \hline
\multicolumn{1}{l|}{\multirow{2}{*}{Model}}   & \multicolumn{1}{c}{${R^2}_c$}         & \multicolumn{1}{c|}{P}             & \multicolumn{1}{c}{${R^2}_c$}         & \multicolumn{1}{c|}{P}             & \multicolumn{1}{c}{${R^2}_c$}         & \multicolumn{1}{c|}{P}             & \multicolumn{1}{c}{${R^2}_c$}         & \multicolumn{1}{c|}{P}             & \multicolumn{1}{c}{${R^2}_c$}         & \multicolumn{1}{c|}{P}             & \multicolumn{1}{c}{${R^2}_c$}         & \multicolumn{1}{c|}{P}             \\
\multicolumn{1}{l|}{}                         & \multicolumn{1}{c}{\textbf{0.05}} & \multicolumn{1}{c|}{\textbf{0.00}} & \multicolumn{1}{c}{\textbf{0.02}} & \multicolumn{1}{c|}{\textbf{0.00}} & \multicolumn{1}{c}{\textbf{0.06}} & \multicolumn{1}{c|}{\textbf{0.00}} & \multicolumn{1}{c}{\textbf{0.07}} & \multicolumn{1}{c|}{\textbf{0.00}} & \multicolumn{1}{c}{\textbf{0.01}} & \multicolumn{1}{c|}{\textbf{0.02}} & \multicolumn{1}{c}{\textbf{0.05}} & \multicolumn{1}{c|}{\textbf{0.00}}
\end{tabular}}

\label{tab:regression_results}
\end{table}

\section{Discussion}

The developed pipeline was able to detect a significant systematic topical gender bias that presents searches for female German politicians with less suggestions on lower average ranks that can be associated with politics and economics. Similarly, the findings show a topical age bias. Query suggestions for older politicians consist of fewer and lower-ranked suggestions associated with the cluster of personal topics and more and higher-ranked suggestions that fit the politics and economics cluster. The overall percentage of explained variance in the metrics seems low, but without comparison and assuming that many unknown factors influence the topics of query suggestions, the results are satisfactory. It seems that the quality of the identified clusters is essential for the effectiveness of the bias identification abilities and the insights the pipeline can produce. By introducing more carefully selected groups of query suggestions as topical clusters, possibly by full or partial manual selection of topic words or utilizing a language model based methodology, the bias identification capabilities could be enhanced further. Another subject for a follow-up study is to test how the pipeline performs on non-person-related searches.

DCG has shown to be a useful metric for describing the perceived topical affiliation but can only be interpreted relative to other DCG scores. It can therefore be used to describe systematic topical bias. The nDCG score can describe the average rank of a cluster or single suggestion. This leads to results similar to the DCG scores if the percentages of terms of the clusters are comparable. For rare or single terms, or if the cluster sizes differ greatly, the metric might be a very useful measure. This could not be tested with the used dataset, however. Compared to the simple percentages of cluster words, the ranking-aware metrics DCG and nDCG revealed more bias. Since the rank- and frequency-aware metrics offer more insight without compromising effectiveness, this speaks in favor of the introduced metrics. Directly comparing the new metrics to the old metric is difficult because the primary defining attribute of the perception-aware metrics is that a different kind of bias is measured. The ability of the pipeline to reveal bias was enhanced by introducing the perception-aware metrics. The results by Bonart et al. explained a little more of the variance inherent in the used metric in cluster 3. However, more significant topical bias was discovered towards more of the groups of meta-attributes and in more of the topical clusters. The new pipeline showed significant biases for clusters 1 and 2 and identified systematic topical biases towards the age and gender and some of the parties and federated states. Overall, the findings and the number of groups in which bias was discovered suggest an improvement to the bias detection capabilities. Due to the biased nature of language and mental representations, natural language data inherits bias as well. For this reason, minimizing the perceptible bias might be more reasonable than trying to completely debias data.

The developed metrics are applicable to any other topical bias analysis of ranked lists, for example, online search results. However, the pipeline does not scale well to other domains. Tests have shown the pipeline's application in the analysis of non-person-related search to be less effective. Search queries and query suggestions beginning with politicians' names seem to follow limited patterns. They consist primarily of the name and a single keyword or a few keywords that show a precise information need. As the produced clusters demonstrate, these terms further fit a limited number of clusters considerably well. Since the quality and precision of the defined topical clusters are critical for the topical bias analysis, the quality of the produced clusters is essential for the bias detection capabilities. Brief tests on a collection of more diverse search queries have shown the pipeline's performance deteriorating. Therefore, to make the pipeline more universally applicable, the clustering needs to be reworked. Since this work focuses on introducing more effective metrics, reworking this drawback of the pipeline has not been addressed. Nevertheless, future work could solve this issue by introducing a more effective topical clustering methodology.

\section{Conclusion}
The main goal was to introduce perception-aware metrics for bias detection in query suggestions of person-related searches. Integrating rank and frequency of cluster words into the bias detection pipeline enables detecting bias that considers how query suggestions are perceived. This is achieved by adopting the DCG and nDCG metrics for bias detection.

By combining perception-aware metrics with topical clustering of query suggestions, the bias detection pipeline is able to overcome the challenges posed by the sparse character of query suggestions. The results presented in section~\ref{results} are more meaningful and better interpretable than the results produced by the pipeline by Bonart et al. Perception-aware bias metrics represent a novel approach to bias detection in query suggestions that could prove useful for other bias detection scenarios as well. 

\pagebreak
\bibliographystyle{splncs04}
\bibliography{lib}

%\noindent\\
%The validity of all links was checked on 07/01/2021.

\end{document}